\documentclass[prl,aps,floats,superscriptaddress,floatfix,twocolumn,longbibliography]{revtex4-1}
\usepackage{amssymb,amsmath}
\usepackage{amsmath,amssymb}
\usepackage{enumitem,kantlipsum}
\usepackage{lipsum}
\usepackage{soul}
\usepackage{cancel}
\usepackage{gensymb}
\usepackage{graphicx}
\usepackage{psfrag}
\usepackage{color}
\usepackage[dvipsnames]{xcolor}
\usepackage{dcolumn}
\usepackage{bm}
\usepackage[normalem]{ulem}
\usepackage{comment}

\usepackage{natbib}
\usepackage{hyperref}
\hypersetup{
  colorlinks=true,
  citecolor=blue,
  linkcolor=blue,
  urlcolor=blue
}

\newcommand{\grad}{\bm\nabla}
\newcommand{\qvec}{\bm q}

\newcommand{\dd}{\mathrm d}

\begin{document}
\title{
{Activity-induced  flatness, instabilities and pattern formation in fluid membranes}}
\author{Debayan Jana}\email{debayanjana96@gmail.com}
\affiliation{Theoretical Physics Division, Saha Institute of Nuclear Physics, a CI of Homi Bhabha National Institute, 1/AF, Bidhannagar, Calcutta 700064, West Bengal, India}
\author{Astik Haldar}\email{astik.haldar@gmail.com}
\affiliation{ Center for Biophysics \& Department of Theoretical Physics, Saarland University, 66123 Saarbr\"ucken, Germany}
\author{Abhik Basu}\email{abhik.123@gmail.com, abhik.basu@saha.ac.in}
\affiliation{Theoretical Physics Division, Saha Institute of Nuclear Physics, a CI of Homi Bhabha National Institute, 1/AF, Bidhannagar, Calcutta 700064, West Bengal, India}

\begin{abstract}
We show that microscopically inversion-asymmetric, permeable active fluid membranes are statistically flat and effectively inversion-symmetric at large scales. Their fluctuations are governed by an asymptotically exact linear hydrodynamic equation  giving orientational long-range order and positional quasi-long-range order.
At intermediate scales, their dynamics is described by a Kardar-Parisi-Zhang equation with spatially long-range noise, giving rough membranes with short-range translational and long-range orientational order described by exactly known scaling exponents. Active stresses can destabilize sufficiently large membranes at the longest scales, while sufficiently strong active permeation flow can drive finite-wavevector instabilities and pattern formation having a nonmonotonic dependence on the strength of activity. Our theory reveals novel activity-driven routes to destabilization and pattern formation in fluid membranes. 

\end{abstract}

\maketitle

Equilibrium membranes are now well understood within continuum theories, where their properties are characterized by a set of elastic constants~\cite{mem1,mem2,mem3}. Living cell membranes, however, display qualitatively different dynamical fluctuations arising from their nonequilibrium nature, driven by microscopic active processes such as cytoskeletal activity and active membrane proteins~\cite{bruce,mizuno,huang1,huang2}. The generality of these active phenomena has motivated the development of minimal theoretical models of biological membranes~\cite{dob1,chen,allard,takatori,bruce,sriram-rev,joanny-prost,sriram-RMP}. Indeed, macroscopic properties of a lipid membrane depend crucially on whether it is active~\cite{tirtha-NJP} or passive~\cite{tirtha-PRE}.

\begin{figure}[b]
\includegraphics[width=0.45\textwidth]{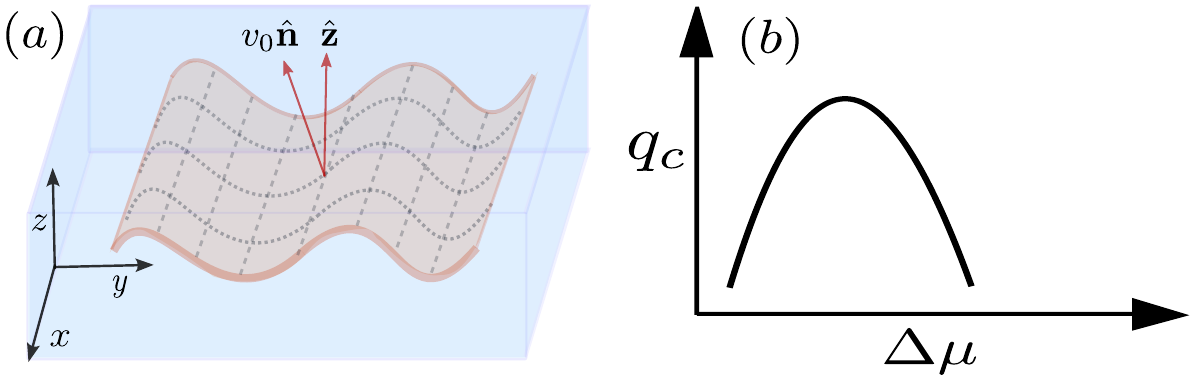}
\caption{ Schematic diagram of (a) the membrane suspended in a 3D bulk isotropic fluid. (b) the variation of $q_c$ with $\Delta \mu$ in the patterned state. See text.}
\label{basic-fig}
\end{figure}

In this Letter, we show how the nonequilibrium dynamics of nearly flat, inversion-asymmetric active membranes arise when coupled to a three-dimensional (3D) active isotropic fluid; see Fig.~\ref{basic-fig}(a) for a schematic model diagram. Our principal results are as follows. (i) When stable, an active anisotropic fluid membrane is statistically flat at long wavelengths, with orientational long-range order and positional quasi-long-range order controlled by a positive active hydrodynamic tension. These results are exact asymptotically, as all nonlinearities are irrelevant in the renormalization group (RG) sense, and inversion symmetry appears as an emergent symmetry. (ii) With stabilizing active permeation, the intermediate-scale dynamics is governed by a Kardar-Parisi-Zhang (KPZ) equation with spatially long-range noise, for which exact scaling exponents are known; inversion asymmetry is retained at these scales. (iii) Strong destabilizing permeation flows render undulation modes unstable at intermediate scales, producing steady patterns.  However, these patterns can be suppressed by the nonlinear effects. The pattern forming instability is argued to depend nonmonotonically on the rate of energy consumption per unit mass $\Delta\mu$; see a schematic diagram of the pattern preferred wavevector $q_c$ versus $\Delta\mu$ in Fig.~\ref{basic-fig}(b).  (iv) A negative active hydrodynamic tension destabilizes the membrane at the largest scales, possibly signalling crumpling. Our theory generalizes Ref.~\cite{sm-ab} to inversion-asymmetric active fluid membranes and demonstrates the crucial role of activity coupled with inversion asymmetry and nonlinearities in the resulting nonequilibrium dynamics.

Below we construct the hydrodynamic equations for a nearly flat patch of a permeable active inversion-asymmetric fluid membrane~\cite{cai} that is embedded in a 3D bulk isotropic fluid. We use the Monge gauge, described by a single-valued height field $h({\bf x},t)$ at a height $z=h(x,y)$ above an arbitrarily located reference plane along the $xy$ plane.
 The dynamics of $h$ is governed by
\begin{equation}
 \frac{\partial h}{\partial t} - v_\text{perm}=v^\prime_z(z=h),\label{eqh}
\end{equation}
where $v_\text{perm}$ gives the contribution from the permeation velocity locally normal to the membrane to the change in $h$ in unit time that in general depends on the local conformation $h$ of the membrane in a tilt-invariant way. In general to the lowest order in fluctuations then
\begin{align}
v_\text{perm}&=(v_0+\nu_{ij}\partial_i\partial_j h  -\mu_p\frac{\delta {\cal F}}{\delta h})\sqrt{1+({\boldsymbol\nabla} h)^2}\nonumber \\&\approx v_0+\nu_{ij}\partial_i\partial_j h -\mu_p\frac{\delta {\cal F}}{\delta h} +  \frac{\lambda}{2}({\boldsymbol\nabla}h)^2,\label{vperm}
\end{align}
where $v_0=\lambda$, a constant, sets the scale of drift of the membrane, $\mu_p$ is a permeation coefficient, $\nu_{ij}$ is a non-hydrodynamic tension, ${\cal F}=\int d^2x [\frac{K}{2} (\nabla^2 h)^2 +C\nabla^2 h]$ is the free energy of a tensionless equilibrium fluid membrane; $K>0$ is the bending stiffness and $C$ is the spontaneous curvature, $\nabla^2 \equiv \partial_x^2 + \partial_y^2$. The square root factor gives the change in $h$ (measured along the $z$ axis) along the local normal~\cite{stanley,debayan-aniso-kpz} due to a local membrane movement, and the last line in (\ref{vperm}) holds for small fluctuations. Furthermore, $v^\prime_z$ is the $z$ component of the (3D) hydrodynamic flow velocity $v_\alpha^\prime$ of the embedding active isotropic bulk fluid. Neglecting inertia in the extreme Stokesian limit, $v_\alpha^\prime$ can be solved from the force balance condition $\nabla_\beta\sigma^\prime_{\alpha\beta}=0;\, \alpha,\beta=x,y,z$, where stress tensor $\sigma_{\alpha\beta}^\prime$ is given by
\begin{equation}
 \sigma^\prime_{\alpha\beta}=\eta'(\partial_\alpha v'_\beta+\partial_\beta v'_\alpha)-P^\prime\delta_{\alpha\beta}
\end{equation}
for an incompressible bulk fluid~\cite{landau-fluid}.  See the Supplemental Material (SM)~\cite{sm} for calculational details. Here, $\eta'$ is the shear viscosity of the bulk fluid and $P'$ is the 3D pressure.

We define a two-dimensional (2D) anisotropic active stress tensor \(\sigma_{\alpha\beta}^a\)
\begin{equation}
 \sigma^a_{\alpha \beta}=\zeta_{\alpha \beta \gamma\delta}  n_\gamma n_\delta,\label{active-stress}
\end{equation}
where $\zeta_{\alpha \beta \gamma\delta}$ is a phenomenological tensor that encodes the anisotropy of the membrane and ${\bf n}=(\hat{\bm{z}},-\nabla_i h)$ is the local normal to the leading order in $h$ fluctuations with $\hat{\bm{z}}$ being
the unit vector along the $z$ direction. This active stress can
locally generate forces $\nabla_j \sigma^a_{z j}$ normally
to the membrane. It can potentially arise, e.g., in a membrane with
actin filaments grafted normally to it (or also in a isotropic live bacterial suspension).  Now define a polarization vector $ {\bf p}$ that is non-vanishing only at the location of the membrane. Elsewhere in the bulk active fluid being in its isotropic phase, we set $ {\bf p}=0$.  For a nearly flat membrane, the condition of normal anchoring gives ${\bf p} = -{\bf n}$~\cite{niladri-epje3}. A non-zero $p_i$ on the membrane allows us to define an anisotropic active stress $\sim\zeta_{\alpha\beta\gamma\delta} p_\gamma   p_\delta$~\cite{sriram-RMP}, which with normal anchoring gives $\sigma^a_{\alpha\beta}$ as defined above.

By using the normal stress balance condition at the membrane, no slip boundary condition and defining $\zeta_1\equiv \zeta_{zxzx}+\zeta_{zxxz}, \zeta_2 \equiv \zeta_{zyzy}+\zeta_{zyyz}$  assuming invariance under $x\rightarrow -x, y\rightarrow -y$, separately, we get in Fourier space
\begin{eqnarray}
-4\eta'q v_z &=&-[\zeta_1q_x^2  +\zeta_2 q_y^2]h({\bf q},t)+\frac{\delta {\cal F}}{\delta h},
\end{eqnarray}
see SM~\cite{sm}. Now using (\ref{eqh}) together with (\ref{vperm}), we get
\begin{eqnarray}
&&\frac{\partial h}{\partial t}
=
\frac{\zeta_1q_x^2+\zeta_2q_y^2}{4\eta' q}h({\bf q},t)
-\frac{\delta{\cal F}}{\delta h}
\left(\frac{1}{4\eta'q}+\mu_p\right)\nonumber \\&&
-(\nu_1q_x^2+\nu_2q_y^2)h
-\frac{\lambda}{2}\sum_{\bf q}
{\boldsymbol q}\cdot({\bf k-q})
h_{\bf q}h_{\bf k-q}
+g_h,
\label{basic-eq}
\end{eqnarray}
where $\nu_{xx}=\nu_1,\nu_{yy}=\nu_2,\nu_{xy}=0=\nu_{yx}$,  $v_0$ has been absorbed by going to a comoving frame; we have added a noise $g_h$ to describe fluctuations.
Equation~(\ref{basic-eq}) is the hydrodynamic equation for an active inversion-asymmetric anisotropic fluid membrane embedded in a 3D isotropic fluid, and forms the principal result from this work. Parameters $\zeta_1,\zeta_2,\nu_1,\nu_2$ are all active coefficients and can be positive or negative. 
The nonlinear $\lambda$ term in (\ref{basic-eq}) violates the inversion symmetry. Thus inversion asymmetry in the hydrodynamic equation is of {\em nonlinear origin}. The noise has two sources, thermal and active noises.
The thermal (equilibrium) noises have a variance~\cite{brochard},  controlled by the Fluctuation-Dissipation-Theorem~\cite{chaikin},
\begin{equation}
 \langle g_h({\bf q},t)g_h({\bf -q},0)\rangle = \frac{2D_h\delta(t)}{4\eta'q},\label{noiseh}
\end{equation}
where $D_h>0$ scales with temperature. The active noises are expected to be short range, and hence subdominant to the thermal noises in the long wavelength limit, in spite of the model being active~\cite{sm-ab}. 
Linear stability of (\ref{basic-eq}) at the largest scales requires that both $\zeta_1,\zeta_2< 0$. At intermediate ranges of $q$, the anisotropic nonhydrodynamic surface tension terms at ${\cal O}(q^2)$ are important. These terms are local in space. At even higher order, at ${\cal O}(q^3)$, the bend modulus term in ${\cal F}$ contributes $-Kq^3h({\bf q},t)/(4\eta')$ in (\ref{basic-eq}), which is obviously stabilizing. Combining up to the ${\cal O}(q^3)$ linear terms in (\ref{basic-eq}), we can define an effective anisotropic damping
\begin{equation}
\lambda_h({\bf q})\equiv -\frac{\zeta_1q_x^2  +\zeta_2 q_y^2}{4\eta' q} +(\nu_1q_x^2 +\nu_2 q_y^2)+\frac{Kq^3}{4\eta'}.\label{lamh}
\end{equation}

\begin{figure*}[]
\includegraphics[width=0.99	\textwidth]{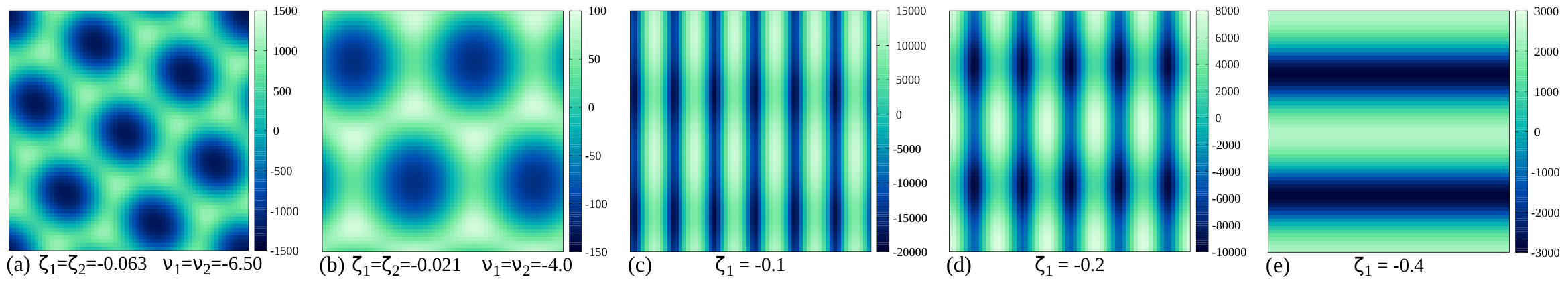}
\caption{Snapshots of the steady-state height profiles (\(L=64\)) showing isotropic and anisotropic pattern with \(\eta'=0.02\), \(K=1\), \(D_h=0.001\), and \(\delta t=0.01\). Panel (a) \(q_c^x=q_c^y\approx0.26\) with \(\lambda=-0.005\), while panel (b) \(q_c^x=q_c^y\approx0.17\) with \(\lambda=-0.05\). The change in \(q_c\) modifies the pattern periodicity. (c)-(e) Anisotropic patterns for different values of \(\zeta_1\), with parameters \(\nu_1=-12\), \(\nu_2=-4\), \(\zeta_2=-0.02\) and \(\lambda=0.001\) (see text).}
\label{iso_diag}
\end{figure*}

In the linearly stable case with $\zeta_1,\zeta_2<0$ and $\nu_1,\nu_2>0$, damping in the long wavelength limit $q^{-1}\gg \xi\sim \nu\eta'/\zeta$, where $\nu=\text{max}(\nu_1,\nu_2),\zeta=\text{min}(|\zeta_1|,|\zeta_2|)$, is controlled by $\zeta_1,\zeta_2<0$:
\begin{equation}
 \lambda_h({\bf q})\approx \frac{|\zeta_1| q_x^2  +|\zeta_2| q_y^2}{4\eta' q}.\label{eff-lamb}
\end{equation}
This immediately gives dynamic exponent $\tilde z=1$ at the largest scales much larger than $\xi$.
The correlation function of $h$ in this stable case can be calculated  from linearized (\ref{basic-eq}) together with the effective damping (\ref{eff-lamb}) and noise variance (\ref{noiseh}). We get
\begin{equation}
 \langle |h({\bf q},t)|^2\rangle = \frac{k_BT}{|\zeta_1| q_x^2 + |\zeta_2| q_y^2}.\label{corr-h}
\end{equation}
The above result is reminiscent of the known equilibrium result for a fluid membrane with a finite anisotropic tension. Nonetheless, both $\zeta_1$ and $\zeta_2$ are {\em dynamical} coefficients of {\em active} origin. Thus in spite of the apparent equilibrium-like behavior, the underlying dynamics is actually of nonequilibrium origin. While the membrane remains anisotropic as in general $\zeta_1\neq \zeta_2$, one can rescale $x$ and $y$ to  bring it to an effectively isotropic form:
 $\langle |h({\bf q},t)|^2\rangle = {k_BT}/(\zeta q^2), \zeta>0.$ 
This in turn gives implies 
\begin{eqnarray}
   \langle [h({\bf r},t)-h(0,t)]^2\rangle \sim \ln \frac{r}{a_0},{\cal W} \equiv \sqrt{\langle (\Delta h)^2}\rangle \sim \sqrt{\ln\left(\frac{ L}{a_0}\right)},\nonumber
 \end{eqnarray}
for $r,L\gg\xi$; $a_0$ is a microscopic cutoff, $\Delta h = h({\bf r},t)-\bar h(t)$ with $\bar h(t)$ being the average height at time $t$. Therefore, $\chi=0$ for $r\gg \xi$ in the linearized limit. Dimensional analysis with  the linear theory exponents $\tilde z=1,\chi=0$ reveals that the $\lambda$ term, which is the most dominant nonlinear term (see below) is {\em irrelevant} in the RG sense in 2D. This in turn means $\tilde z=1,\chi=0$ are {\em exact} in the asymptotic long wavelength limit;
see also Ref.~\cite{sm-ab} for similar results for a symmetric, isotropic membrane. This further means that at these largest scales, inversion symmetry appears as an emerging symmetry at these scales.

To explore scaling at intermediate-range wavevectors $q$ with $q^{-1}\ll\xi$, consider the linearly stable case, $\zeta_1,\zeta_2<0$ and $\nu_1,\nu_2>0$, where the latter set the dominant damping. Assuming sufficiently large scale separation and neglecting the nonlinear $\lambda$ term, Eq.~(\ref{basic-eq}) becomes an anisotropic Edwards-Wilkinson equation for surface growth~\cite{ew} with a long-range noise for $q$ with $q^{-1}\ll \xi$. In this regime, the $\lambda$ term is actually {\em relevant} in 2D. Indeed, dropping the $\zeta_1,\zeta_2$ terms, (\ref{basic-eq}) becomes an anisotropic version of the  Kardar-Parisi-Zhang (KPZ) equation with a long-range noise~\cite{medina}. Galilean invariance of the equation means $\lambda$ does not renormalize~\cite{stanley}. Furthermore, due to the long range nature of the noise, there are no relevant fluctuation-corrections to $D_h$~\cite{medina}. The scaling exponents $\chi$ and $\tilde z$, valid in the intermediate scales, can be obtained {\em exactly} by using the nonrenormalization of $\lambda$ and noise strength $D_h$. This gives
\begin{eqnarray}
 \chi&=&\frac{1}{3},\,\tilde z=\frac{5}{3}.
 \end{eqnarray}
 Since $0<\chi<1$, the surface is rough. In particular,
 \begin{equation}
  \langle [h({\bf r},t)-h(0,t)]^2\rangle \sim r^{2/3},\,{\cal W} \sim L^{1/3},
 \end{equation}
for $r,L\ll\xi$.
We remind the reader that the above results are only valid on intermediate length scales, where $\nu_1,\nu_2>0$ dominate over $\zeta_1,\zeta_2$. 

We now discuss the linear instabilities in the model. Considering the linearized limit of (\ref{basic-eq}) when one or both $\zeta_1,\zeta_2>0$, or one or both $\nu_1,\nu_2<0$, the model is unstable at the largest or intermediate scales.  Notice the formal similarities between linearly unstable Eq.~(\ref{basic-eq}) and the Kuramoto-Sivashinsky equation for flame front propagation~\cite{kura1,kura2,kura3}, whose  long wavelength fluctuations belong to the  2D KPZ equation~\cite{rahul-jaya}.

\begin{figure}[h]
\includegraphics[width=0.49\textwidth]{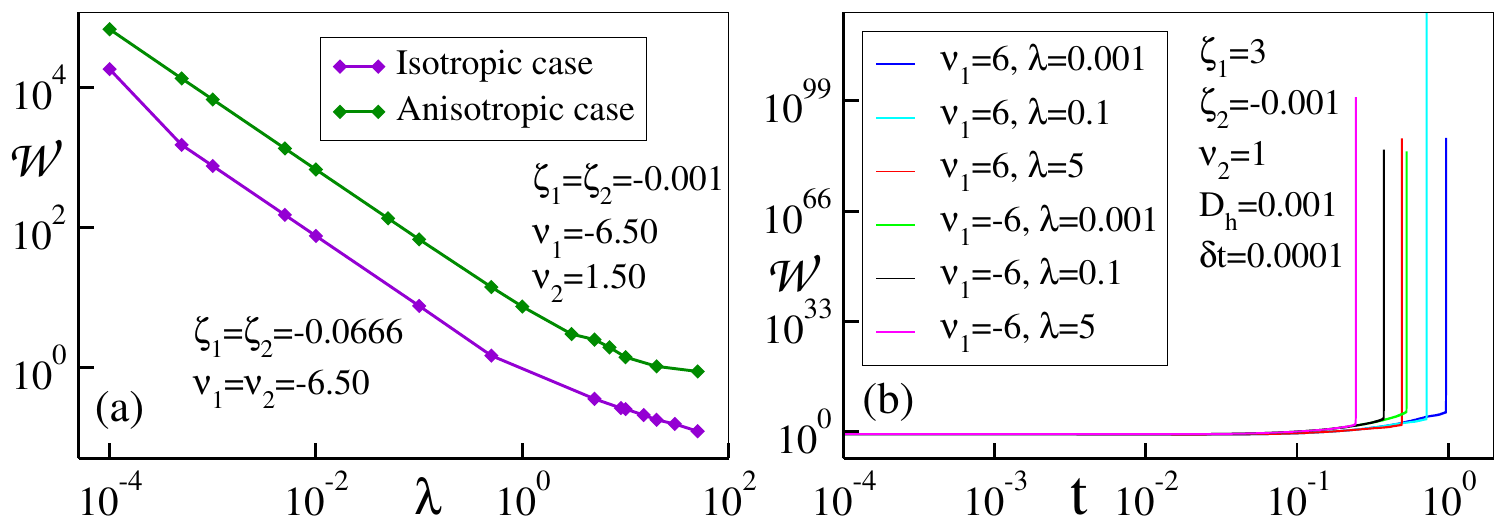}
\caption{ Log-log plots of (a) \(\mathcal{W}\) versus \(\lambda\) for the isotropic case (\(D_h=0.001\)) and the anisotropic case (\(D_h=0.1\)) with \(\delta t=0.001\). The height fluctuations decrease with increasing \(|\lambda|\). (b) \(\mathcal{W}\) versus \(t\) for different values of \(\nu\) and \(\lambda\). When at least one of the \(\zeta_i\) terms is destabilizing, \(\mathcal{W}\) grows rapidly, and diverges, possibly indicating membrane crumpling. Other parameters: \(\eta'=0.02\), \(K=1\) and $L=64$, (see text).}
\label{crumple_diag}
\end{figure}

For high enough $q$, the stabilizing $K q^3/(4\eta' )$ term in \eqref{basic-eq} dominates. From (\ref{lamh}), $\lambda_h$ should be extremum at some ${\bf q}_c=(q^x_{c},q^y_{c})$, which is the preferred or selected wavevector.
The threshold of linear instability and ${\bf q}_c$ are determined by setting $q_x=q\cos\theta,\,q_y=q\sin\theta$ in Eq.~(\ref{lamh}) and imposing the conditions $\left.\partial \lambda_h/\partial q\right|_{q=q_c,\theta=\theta_c}=0$, 
and 
$\left.\partial \lambda_h/\partial \theta\right|_{q=q_c,\theta=\theta_c}=0$. See also SM~\cite{sm}. These yield
\begin{equation}
q_c = \frac{2\eta'}{3K}\left(-2\nu(\theta_c)+ \sqrt{4\nu^2(\theta_c)+\frac{3K}{4\eta'^2}\zeta(\theta_c)}\right),\label{qc-soln}
\end{equation}
together with the condition $\sin 2\theta_c = 0$, giving $\theta_c=0,\pi/2$. Here $\nu(\theta)=\nu_1\cos^2\theta+\nu_2\sin^2\theta$ and $\zeta(\theta)=\zeta_1\cos^2\theta+\zeta_2\sin^2\theta$. In \eqref{qc-soln}, only the larger solution for $q_c$ is written, assuming a growth rate $\exp[\nu(\theta_c)q^2_c(\theta_c)t]$, which is dominated by the larger solution of $q_c$.
%
This gives two possible values of the preferred wavevector,
\begin{align}
&q^a_c = \frac{2\eta'}{3K}\left(-2\nu_b+\sqrt{4\nu_b^2+\frac{3K}{4{\eta'}^2}\zeta_b}\right), \label{q-a-c} 
\end{align}
$a=x,b=1; a=y,b=2$. When $q_c^x\neq q_c^y$, assuming a growth rate $\exp[\nu(\theta_c)q^2_c(\theta_c)t]$, the linear instability is controlled by the larger among $q_c^x,q_c^y$. From~\eqref{q-a-c}, $\nu_b$ must exceed a finite threshold $(\nu_b^c)^2= -3K\zeta_b/(16{\eta'}^2)$ for a pattern forming linear instability to set in.
{Linear stability analysis, thus, suggests that if any of $\zeta_1,\zeta_2,\nu_1,\nu_2$ flips sign from its required sign in the linearly stable case, a nonzero $q^x_{c},q^y_{c}$ can be possible, suggesting formation of patterns. We use direct numerical simulations (DNS) of  Eq.~(\ref{basic-eq}) using a pseudo-spectral method with the \(2/3\) de-aliasing rule to study the patterns; see SM~\cite{sm} for technical details.
In the isotropic case with \(\zeta_1=\zeta_2<0\) and \(\nu_1=\nu_2<0\), gives $q_c^x=q_c^y$ with $\theta_c$ undetermined. Figs.~\ref{iso_diag}(a) and~\ref{iso_diag}(b) display isotropic mound-like patterned states with no preferred growth direction. The patterns are characterized by nearly symmetric spatial structures along both \(x\) and \(y\) directions. As \(q_c\) decreases from \(0.26\) to \(0.17\), the pattern periodicity increases, leading to larger characteristic structures. In the anisotropic case with \(\nu_1,\nu_2<0\) and \(\zeta_1,\zeta_2<0\), with \(\nu_1\neq\nu_2\) and \(\zeta_1\neq\zeta_2\), the pattern is controlled by the dominant $q_c$. The anisotropic patterns in Fig.~\ref{iso_diag}(c,d,e) arise from the competition between the destabilizing $\nu_i$ and stabilizing $\zeta_i$ terms. To understand this we fix $q_c^y\approx0.17$, and tune $q_c^x$ by varying $\zeta_1$. For $\zeta_1=-0.1$, the dominant mode $q_c^x\approx0.57$ leads to stripes parallel to the $y$ axis. Increasing $|\zeta_1|$ reduces $q_c^x$, thereby increasing the spatial periodicity. For $\zeta_1=-0.4$, the instability along the $x$ direction is completely suppressed, and the surviving mode $q_c^y\approx0.17$ leads to stripes parallel to the $x$ axis. Thus, varying $\zeta_1$ controls both the periodicity and orientation of the stripes. See SM~\cite{sm} for additional details.

For a sufficiently large $|\lambda|$, the patterns due to $\nu_1$ and/or $\nu_2$ disappear~\cite{sm}. Figure~\ref{crumple_diag}(a) shows variation in the pattern intensity described by ${\cal W}$ with $\lambda$ in steady state. As \(|\lambda|\) increases, the height fluctuations are significantly suppressed; see SM~\cite{sm}. In contrast, when the linear instability is due to $\zeta_1$ and/or $\zeta_2$,  the membrane gets unstable with the fluctuations growing rapidly without any steady pattern and unaffected by $\lambda$. Figure~\ref{crumple_diag}(b) shows the time evolution of the width, \({\mathcal W}\), for a system that is stable along the \(y\) direction but unstable along the \(x\) direction due to the \(\zeta_1\) term. The rapid divergence of \({\mathcal W}\) for different values of \(\lambda\) indicates possible membrane crumpling, whose detailed study is beyond the scope of the present work.
Physically, that the linear instability due to $\nu_1,\nu_2$ ultimately leads to steady patterns, which get progressively suppressed by increasingly stronger nonlinear effects, only means that the corrections to $\nu_1,\nu_2$ due to the nonlinear effects are positive and eventually turn effective $\nu_1,\nu_2$ positive in the long time limit, thereby arresting the instability. A full calculation is beyond the scope of the present work. Instead, we use a workaround to show that the corrections to $\nu_1,\nu_2$ due to the $\lambda$ nonlinearity are positive definite and grow with $|\lambda|$, which strongly suggests possible nonlinearity-induced arrest of the linear instabilities.
To that end, 
we use a bare perturbation theory in which the wavevector integrals in the corrections to $\zeta_1,\zeta_2,\nu_1,\nu_2$ are extended down to a lower cutoff \(q_{\min}=2\pi/\tilde \xi\), where 
$\tilde \xi$ is a ``large scale'', whose precise value is unimportant to the present discussion. In order to have a stable ``unperturbed'' equation about which we expand in powers of $\lambda$, we only keep the stabilizing free energy contribution $-Kq^3/(4\eta')$ in linearized \eqref{basic-eq}, or equivalently, in \eqref{lamh}, and find corrections to the possible destabilizing linear $\zeta_1,\zeta_2,\nu_1,\nu_2$ terms. Now notice that due to the structure of the $\lambda$ term, in which every factor of $h$ comes with a $\boldsymbol \nabla$, the lowest order correction to the propagator must be analytic and at least bilinear in $k_x,k_y$ to all order in perturbation theory. This means there are no corrections to $\zeta_1,\zeta_2$, a fact that can also be argued by stating that the $\zeta_1,\zeta_2$ terms being proportional to $|k|$ are nonanalytic, receive no perturbative corrections from the $\lambda$ term. Thus any instability due to $\zeta_1,\zeta_2>0$ {\em cannot} be suppressed by the $\lambda$ term, for any sign and magnitude of $\lambda$, in a large enough system. Our DNS studies confirm this. In contrast, the  corrections to $\nu_1,\nu_2$ are {\em positive definite}. We find for $\nu^\text{eff}_i$, the effective values of $\nu_i$}
\begin{eqnarray}
 &&\nu^\text{eff}_i \approx \nu_i +\frac{\lambda^2 D_h \eta' }{48\pi^4 K^2}\Bigl(\tilde\xi^3-a^3\Bigl) + {\cal O}(\lambda^4) \label{nu-eff}
\end{eqnarray}
to the lowest order in $\lambda$;
see SM~\cite{sm} for intermediate calculational details.
Equation~(\ref{nu-eff}) implies that, although the bare coefficients \(\nu_i\) are negative, the nonlinear \(\lambda\) term generates positive corrections at long wavelengths, thereby suppressing the instability and stabilizing the system into a patterned phase. Since $\lambda$ scales with $v_0$, the corrections to $\nu_i$ must scale with $v_0$ as $v_0^2$. Since both drift $v_0$ and active tension $\nu_i$ should scale with $\Delta\mu$, we have
\begin{equation}
 \nu^\text{eff}_i = -a\Delta\mu + b\Delta\mu^2 + {\cal O}(\Delta\mu^4),\;a,b>0,\label{effect-nu}
\end{equation}
$a,b\sim {\cal O}(1)$ are dimensional constants parametrized by the model parameters. This means for larger activity, $\nu^\text{eff}_i$ turns positive, and hence the pattern disappears, indicating a {\em nonmonotonic} dependence on $\Delta\mu$. If terms of order higher than ${\cal O}(\Delta\mu^2)$ are included in \eqref{effect-nu}, more complex nonmonotonic behavior can be possible, although our DNS studies have not detected any such behavior.

In the isotropic limit, a linear stability diagram in the $\zeta$-$\nu$ plane should have the same topology as the one in Ref.~\cite{sm-ab}. { While numerical estimates of $\zeta_1,\zeta_2,\nu_1,\nu_2$ for anisotropic membranes are not available to our knowledge, assuming weak anisotropy and considering $\zeta_1\sim\zeta_2\sim\zeta$, we have $\zeta/(4\eta')\sim 100\,$nm/min~\cite{svitkina}. Similarly, assuming $\nu_1\sim\nu_2\sim\nu$, typical values $\nu \sim 10^{-17}$ {m$^2$}/{s}~\cite{chen_biomem}. The sign of $\zeta$ can be varied by considering extensile or contractile active stress~\cite{sriram-rev,joanny-prost,sriram-RMP}. The sign of $\nu$ can be made negative
by strong curvature-dependent active fusion processes on the
membrane~\cite{sarosij,solon_lipid}, or by curvature-inducing molecules~\cite{noguchi}; it can also be positive for certain protein pumps in membranes~\cite{prost-pump,noguchi1}. The drift velocity $v_0\sim 1\mu$m/sec.~\cite{pnas}. In typical cellular processes, $\Delta\mu\sim 20 k_BT$~\cite{,abhik_polar_gel}, and hence can dominate over typical thermal fluctuations.}

In addition, to the $\lambda$-term, there can be other nonlinear terms of equilibrium and active origins; see Ref.~\cite{sm-ab}, which are subleading to the $\lambda$ term in the long wavelength limit.  Therefore, the scaling of the correlation functions in the linearly stable case with $\zeta_1,\zeta_2<0$ as reported above are in fact {\em exact} in the asymptotic long wavelength limit, with inversion symmetry appearing as an {\em emergent symmetry}. Indeed, since all the potential nonlinear terms, including those responsible for thermal crumpling~\cite{luca}, are {\em irrelevant} in the largest scales, the theory is {\em effectively linear} in that limit and the nonlinearities play no role in sustaining the long range orientational order of the membrane. This is unlike most ordered active matter systems~\cite{toner-tu1,toner-tu2,astik-xy1,astik-xy2,debayan1},  where nonlinear effects are essential, with rare exceptions being those studied in Refs.~\cite{sm-ab,bottom1,bottom2}. Lastly, for reasons discussed in Ref.~\cite{debayan-aniso-kpz}, spontaneous curvature $C$ should be irrelevant at the largest scales.



In summary, we have formulated the hydrodynamic theory of an active, permeable lipid membrane immersed in a fluid. In the stable regime, active hydrodynamic tension dominates at the largest scales, rendering the dynamics effectively linear and restoring inversion symmetry as an emergent property. The membrane exhibits orientational long-range order and positional quasi-long-range order. At intermediate scales, active permeation flows dominate, leading to KPZ dynamics with long-range noise; inversion asymmetry persists, orientational long-range order survives, but positional order becomes short-ranged.
Activity can destabilize the flat state in two distinct ways. Destabilizing active permeation produces steady finite-wavelength patterns with possible relevance for the finite size structures observed in membranes~\cite{pattern-mem}, whereas destabilizing active stress yields a long-wavelength instability without pattern formation. Nonlinearities can suppress the patterns generated by permeation-driven instabilities, but do not affect the active stress-driven instability. 
 It will be interesting to explore similar effects in other related nonlinear models~\cite{debayan-mct,debayan-ckpz-mbe,mintu-chemo1,mintu-chemo2}. We expect that in vitro experiments on lipid layers  immersed in an active fluid bath, e.g., a solution of live, orientable bacteria or actin filaments in the isotropic phase or in reconstituted membrane-tethered actin cortices~\cite{betz,mayor} could test the key predictions of our theory.

{\em Acknowledgment:-} A.H. thanks Alexander von Humboldt (AvH) Stiftung (Germany) for a postdoctoral fellowship. A.B. thanks Alexander von Humboldt Stiftung (Germany) for partial financial support through their research group linkage programme (2024), and  ANRF (India) for partial financial support through their ARG (MATRICS) programme (file no.: ANRF/ARGM/2025/000461/TS).

\bibliography{hydro.bib}

\clearpage
\onecolumngrid
\begin{center}
\textbf{\LARGE Supplemental Material}
\end{center}
\vspace{0.5cm}

\section{Hydrodynamic Theory of the Active Membrane}

The 3D stress tensor is of a simple incompressible fluid is given by
\begin{equation}
\sigma^\prime_{\alpha\beta}=\eta'(\partial_\alpha v'_\beta+\partial_\beta v'_\alpha)-P^\prime\delta_{\alpha\beta},
\label{eq:bulk-stress}
\end{equation}
where $P'$ is the bulk-fluid pressure, $\eta'$ is the 3D shear viscosity, and $\alpha,\beta=x,y,z$. Here $\bm v'$ is the velocity of the 3D bulk fluid. In the Stokesian limit, force balance gives $\partial_\beta\sigma'_{\alpha\beta}=0$.
Using Eq.~\eqref{eq:bulk-stress} we obtain,
\[
-\partial_\alpha P'
+\eta'\partial_\beta
\left(\partial_\alpha v'_\beta+
\partial_\beta v'_\alpha\right)=0.
\]
For an incompressible fluid, $\partial_\beta v'_\beta=0$, and hence we obtain,
\begin{equation}
{\eta'\nabla_3^2 v'_\alpha-\partial_\alpha P'=0,}
\label{eq:stokes-real}
\end{equation}
where $\nabla_3^2 \equiv \partial_x^2 + \partial_y^2 + \partial_z^2$.

\subsection{Solutions for the bulk fluid velocity and pressure}

We define in-plane Fourier transform of $v_z'$ by
\begin{equation}
\bm v_z(\bm r_\parallel,z,t)
=\int\frac{\dd^2q}{(2\pi)^2}\,
\bm v_z(\qvec,z,t)e^{i\qvec\cdot\bm r_\parallel},
\label{eq:fourier-def}
\end{equation}
where $\bm r_\parallel=(x,y)$. 
Therefore, $v_z'$ satisfies
\begin{equation}
{\eta'(\partial_z^2-q^2)v'_z(\qvec,z,t)
=\partial_zP'(\qvec,z,t).}
\label{eq:z-ode}
\end{equation}
Taking the divergence of Eq.~\eqref{eq:stokes-real} and using incompressibility we get
\[
\eta'\nabla_3^2(\grad\cdot\bm v')-\nabla_3^2P'=-\nabla_3^2P'=0.
\]
After Fourier transformation in the plane, we obtain
\[
(\partial_z^2-q^2)P'=0.
\]
Therefore,
\[
P'=D_1(\qvec,t)e^{-qz}+D_2(\qvec,t)e^{qz}.
\]
Since the fluid extends to $z\to+\infty$ above the membrane and to
$z\to-\infty$ below it, the decaying solutions are
\begin{equation}
P'(\qvec,z,t)=
\begin{cases}
D_1(\qvec,t)e^{-qz},&z>0,\\
D_2(\qvec,t)e^{qz},&z<0.
\end{cases}
\label{eq:pressure-sol}
\end{equation}

For $v_z'$, by using \eqref{eq:pressure-sol} Eq.~\eqref{eq:z-ode} becomes
\[
(\partial_z^2-q^2)v'_z
=-\frac{qD_1}{\eta'}e^{-qz}
+\frac{qD_2}{\eta'}e^{qz}.
\]
The complementary function is a linear combination of $e^{-qz}$ and
$e^{qz}$ and the particular integral is a linear combination of $ze^{-qz}$ and
$ze^{qz}$.
Consequently,
\begin{equation}
v'_z(\qvec,z,t)=
\begin{cases}
(A_1(\qvec,t)+B_1(\qvec,t)z)e^{-qz},&z>0,\\
(A_2(\qvec,t)+B_2(\qvec,t)z)e^{qz},&z<0,
\end{cases}
\label{eq:vz-sol}
\end{equation}
where $B_1,B_2$ are related to the pressure amplitudes by $B_1=D_1/2\eta'$ and $B_2=D_2/2\eta'$. In Fourier space, incompressibility reads $iq_xv'_x+iq_yv'_y+\partial_zv'_z=0$. Since the in-plane components $v_x',v_y'$ are continuous across the interface at $z=h\approx 0$, $\partial_zv'_z$ is also continuous across it:
\begin{equation}
 {\partial_z v_z'}|_{z=h_+}={\partial_z v_z'}|_{z=h_-},
\end{equation}
Continuity of $v_z'$ and ${\partial_z v_z'}$ across the membrane gives
\begin{equation}
 A_1=A_2,\;\;-qA_1+B_1=qA_2+B_2.
\end{equation}

\subsection{Normal force balance}

The total internal stress tensor of the membrane is
\begin{equation}
\sigma_{ij}^{\mathrm{tot}}
=\eta(\partial_i v_j+\partial_j v_i)
+\eta_b(\grad\cdot\bm v)\delta_{ij}
-P\delta_{ij}+\sigma_{ij}^{a}.
\label{eq:membrane-total-stress}
\end{equation}
Here $\bm v$ is the 2D velocity of the membrane fluid, $i$ and $j$ can take values $x$ or $y$, $\eta$ and $\eta_b$ are the 2D shear and bulk viscosities, $P$ is the 2D pressure.
Furthermore, $\sigma_{ij}^{a}$  is the active stress $\sigma_{\alpha\beta}^{a}$ with $\alpha,\beta=i,j=x,y$.  The active stress is given by
\[
\sigma_{\alpha\beta}^{a}
=\zeta_{\alpha\beta\gamma\delta}n_{\gamma}n_{\delta},
\]
where $\alpha,\beta,\gamma,\delta=x,y,z$, and $\zeta_{\alpha\beta\gamma\delta}$ is a phenomenological tensor characterizing the anisotropy of the membrane. The vector $\bm{n}=(\hat{\bm{z}},-\nabla_i h)$ represents the local membrane normal to leading order in height fluctuations.  This active stress can locally generate forces $\nabla_j \sigma^a_{ij}$ tangentially and $\nabla_j \sigma^a_{z j}$ normally
to the membrane.

The normal stress balance condition at the membrane gives
\begin{equation}
\left.\sigma'_{zz}\right|_{h_+}
-\left.\sigma'_{zz}\right|_{h_-}
-\frac{\delta {\cal F}}{\delta h}
+\partial_j\sigma^a_{zj}=0.
\label{eq:normal-balance-general}
\end{equation}
Here ${\cal F}$ is the membrane free energy functional, and
$\partial_j\sigma^a_{zj}$ is the $\hat{\bm z}$-directed force per unit area
arising from the activity due to actin filaments. Thus using Eq.~\eqref{eq:bulk-stress}, Eq.~\eqref{eq:normal-balance-general} becomes
\begin{align}
\left(2\eta'\frac{\partial v'_z}{\partial z}-P'\right)\bigg|_{z=h_+}
-\left(2\eta'\frac{\partial v'_z}{\partial z}-P'\right)\bigg|_{z=h_-}-\frac{\delta {\cal F}}{\delta h}
+\partial_j\sigma^a_{zj}=0.
\label{eq:normal-balance-expanded}
\end{align}
Expanding the last term in the left hand side of (\ref{eq:normal-balance-expanded}), we get
\begin{align}
\partial_j \sigma_{zj}^a
={}&-(\zeta_{zxzx}+\zeta_{zxxz})\partial_x p_x
-(\zeta_{zyzy}+\zeta_{zyyz})\partial_y p_y \nonumber\\
&-(\zeta_{zxzy}+\zeta_{zxyz})\partial_x p_y
-(\zeta_{zyzx}+\zeta_{zyxz})\partial_y p_x \nonumber\\
={}&-(\zeta_1 \partial_x^2 h+\zeta_2 \partial_y^2 h).
\label{active-stress-expand}
\end{align}
where in the last line in (\ref{active-stress-expand}), we have imposed invariance under $x\rightarrow -x$ and $y\rightarrow -y$ separately; $\zeta_1,\zeta_2$ are active coefficients with $\zeta_1\equiv \zeta_{zxzx}+\zeta_{zxxz}, \zeta_2 \equiv \zeta_{zyzy}+\zeta_{zyyz}$. Using the velocity and pressure solutions above, evaluated at $z=0$, the difference of normal stresses reduces to
\begin{equation}
-4\eta' q\,v'_z
=\frac{\delta {\cal F}}{\delta h}
-\partial_j\sigma^a_{zj},
\label{eq:normal-traction-result}
\end{equation}
giving
\begin{equation}
{
v'_z=
\frac{1}{4\eta' q}
\left[-\frac{\delta {\cal F}}{\delta h}
+\partial_j\sigma^a_{zj}\right].
}
\label{eq:vz-at-membrane}
\end{equation}
Substituting the expressions for $v'_z$ from Eq.~\eqref{eq:vz-at-membrane} and $v_{\mathrm{perm}}$ from Eq.~\eqref{vperm} of the main text into Eq.~\eqref{eqh}, we obtain
\begin{align}
\frac{\partial h}{\partial t}
={}
\nu_1\frac{\partial^2h}{\partial x^2}
+\nu_2\frac{\partial^2h}{\partial y^2}
-\mu_p\frac{\delta F}{\delta h}
+\frac{\lambda}{2}(\grad h)^2
+\frac{1}{4\eta' q}
\left[
-\frac{\delta F}{\delta h}
+\partial_j\sigma^a_{zj}
\right].
\label{eq:height-eom-intermediate}
\end{align}
The mixed derivative term in $\nu_{ij}$, proportional to $\partial_x \partial_y h$, is omitted as it is forbidden by the imposed invariance under $x\rightarrow -x$ and $y\rightarrow -y$ separately. Using expression~\eqref{active-stress-expand} in above equation, we obtain,
\begin{align}
\frac{\partial h}{\partial t}
={}
-\frac{
\zeta_1\partial_x^2h+\zeta_2\partial_y^2h
}{4\eta' q}
-\frac{\delta F}{\delta h}
\left(\frac{1}{4\eta' q}+\mu_p\right)
+\nu_1\partial_x^2h+\nu_2\partial_y^2h
+\frac{\lambda}{2}(\grad h)^2+g_h.
\label{eq:height-eom}
\end{align}
In Fourier space this reads as,
\begin{equation}
\frac{\partial h}{\partial t}
={}
\frac{\zeta_1q_x^2+\zeta_2q_y^2}{4\eta' q}h({\bf q},t)
-\frac{\delta{\cal F}}{\delta h}
\left(\frac{1}{4\eta'q}+\mu_p\right)
-(\nu_1q_x^2+\nu_2q_y^2)h
-\frac{\lambda}{2}\sum_{\bf q}
{\boldsymbol q}\cdot({\bf k-q})
h_{\bf q}h_{\bf k-q}
+g_h,
\end{equation}
which is Eq.~(\ref{basic-eq}) of the main text. Here a noise $g_h$ is added to describe fluctuations.

\section{Corrections to $\nu_1,\nu_2$ by the nonlinear $\lambda$ term}


We start by constructing a generating functional~\cite{bausch,tauber} by using Eq.~(\ref{basic-eq}) and Eq.~(\ref{noiseh}) of the main text. We get
\begin{equation}
 \mathcal{Z}=\int \mathcal{D}\hat{h} \mathcal{D}h e^{-\mathcal{S}[\hat{h},h]},
\end{equation}
where $\hat{h}$ is referred as response field and $\mathcal{S}$ is action functional, which is given by
\begin{align}
S= -\int_{{\bf k},t}\frac{D}{|{\bf k}|}|\hat{h}|^2 + \int_{{\bf k},t}\hat{h}\Bigg[\partial_th-\frac{\zeta_1k_x^2  +\zeta_2 k_y^2}{4\eta' |{\bf k}|}h+(\nu_1k_x^2 +\nu_2 k_y^2)h+\kappa |{\bf k|}^3h\Bigg]- \int_{{\bf x},t}\hat{h}\Bigg[\frac{\lambda}{2} {({\boldsymbol\nabla}h)^2}\Bigg].\label{action}
\end{align}
Here $D=\frac{D_h}{4\eta'}$ and $\kappa=\frac{K}{4\eta'}$.


\subsection{Results from the linearized equation of motion}

In this Section, our primary interest is the dependence of the patterned phase on the nonlinear coupling $\lambda$, which exists at intermediate length scales. To ensure a stable ``unperturbed'' state in our perturbation theory, we construct the bare two-point function using only the stabilizing free-energy contribution in the linearized version of Eq.~\eqref{basic-eq} of the main text. We first define the Fourier transform in space and time as
\begin{equation}
h({\bf x},t)=\int_{{\bf k},\omega} h({\bf k},\omega)
e^{i({\bf k}\cdot{\bf x}-\omega t)}.
\end{equation}
Using the Gaussian part of the action functional~(\ref{action}), obtained by setting \(\lambda=0\), and neglecting the lower-order terms proportional to \(\zeta_i\) and \(\nu_i\), ($i=1,2$) the bare two-point functions are,

\begin{subequations}
\begin{align}
&\langle \hat{h}({\bf k},\omega) \hat{h}(-{\bf k},-\omega)\rangle_0=0,\\
&\langle \hat{h}({\bf k},\omega) h(-{\bf k},-\omega)\rangle_0=\frac{1}{i\omega +\kappa |{\bf k}|^3},\\
&\langle \hat{h}(-{\bf k},-\omega) h({\bf k},\omega)\rangle_0=\frac{1}{-i\omega +\kappa |{\bf k}|^3},\\
&\langle h({\bf k},\omega) h(-{\bf k},-\omega)\rangle_0=\frac{2D|{\bf k}|^{-1}}{\omega^2 +\kappa^2|{\bf k}|^6}.
\end{align}
\end{subequations}

\begin{figure}[h]
\centering
\includegraphics[width=0.2\textwidth]{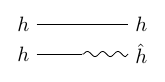}
\caption{Diagrammatic representations of two-point functions.}
\label{propagator}
\end{figure}
Fig.~\ref{propagator} shows the diagrammatic representations of the various  two-point functions.



\subsection{One-loop correction to the model parameter $\nu_i$}

\begin{figure}[t]
\centering
\includegraphics[width=0.4\textwidth]{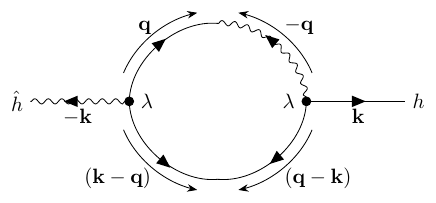}
\caption{One-loop Feynman diagram that contributes to the correction of $\nu_i$.}
\label{nu_diag}
\end{figure}

The one-loop Feynman diagram, with a symmetry factor 8, that contributes to the fluctuation correction of $\nu_i$ is shown in Fig.~\ref{nu_diag}.
It is given by
\begin{align}
\lambda^2\int_{{\bf q},\Omega}\big[{\bf q}\cdot {\bf (k-q)}\big]\big[{\bf k}\cdot {\bf (q-k)}\big]\times\frac{1}{-i\Omega+\kappa |{\bf q}|^3}\times\frac{2D}{|{\bf k-q}|(\Omega^2+\kappa^2|{\bf k-q}|^6)}.
\end{align}
After performing the $\Omega$ integral, we obtain,
\begin{align}
\frac{\lambda^2D}{\kappa^2}k_j\int_{{\bf q}}\frac{q_i(k-q)_i(q-k)_j}{|{\bf k-q}|^4\bigl[|{\bf k-q}|^3+ |{\bf q}|^3\bigl]}.
\end{align}
After symmetrizing, i.e., ${\bf q}\rightarrow {{\bf q}+{{\bf k}}/{2}}$, we are left with two integrals that include corrections for $\nu_i$. These integrals are
\begin{align}
\mathcal {I}_1&=-\frac{\lambda^2D}{\kappa^2}k_jk_m\int_{{\bf q}}\frac{q^2q_jq_m}{|{\bf q}|^9},\\
&=-\frac{\lambda^2D}{\kappa^2d}k^2\int_{{\bf q}}\frac{d^dq}{|{\bf q}|^5},
\end{align}
and
\begin{align}
\mathcal {I}_2&=\frac{\lambda^2D}{4\kappa^2}k^2\int_{{\bf q}}\frac{d^dq}{|{\bf q}|^5}.
\end{align}
Adding $\mathcal {I}_1$ and $\mathcal {I}_2$ we obtain bare perturbative correction to $\nu_i$,
\begin{align}
-k^2\frac{\lambda^2D}{\kappa^2}\frac{4-d}{4d}\int_{{\bf q}}\frac{d^dq}{|{\bf q}|^5}.
\end{align}
Thus the resulting effective coefficient \(\nu_i^e\) is given by,
\begin{align}
 \nu^e_i \approx \nu_i +\frac{\lambda^2D}{\kappa^2}\frac{4-d}{4d}k_d\int^{2\pi/a}_{2\pi/\tilde\xi}\frac{|{\bf q}|^{d-1}}{|{\bf q}|^5}dq.
\end{align}
Here $\tilde \xi$ is a ``large'' length scale whose precise value is unimportant here, $a$ is microscopic cutoff and \(k_d=\frac{S_d}{(2\pi)^d}\), with \(S_d\) being the surface area of a \(d\)-dimensional unit sphere. For \(d=2\), the resulting effective coefficient \(\nu^\text{eff}_i\) is given by,
\begin{eqnarray}
 &&\nu^\text{eff}_i \approx \nu_i +\frac{\lambda^2 D_h \eta' }{48\pi^4 K^2}\Bigl(\tilde\xi^3-a^3\Bigl), \label{nu-eff}
\end{eqnarray}
as mentioned in the main text.

\section{Characteristic pattern wavelength from linear theory}

\begin{figure}[t]
\centering
\includegraphics[width=0.4\textwidth]{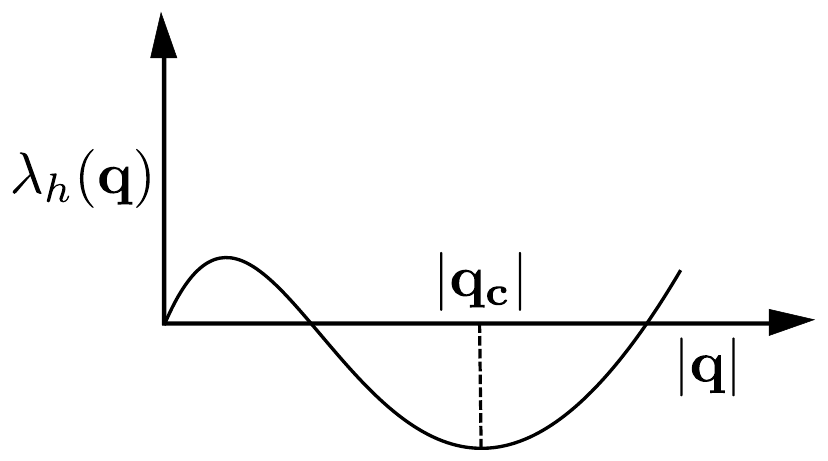}
\caption{Schematic plot of \(\lambda_h({\bf q})\) versus \(|{\bf q}|\). The surface is stable at small and large length scales (\(\lambda_h>0\)) but unstable at intermediate scales (\(\lambda_h<0\)), resulting in the emergence of patterned states.}
\label{lam_vs_q_diag}
\end{figure}

Considering the linear part of Eq.~(\ref{basic-eq}) of the main text, we can write
\[
\frac{\partial h({\bf q},t)}{\partial t}
=
-\lambda_h({\bf q})\,h({\bf q},t),
\]
where
\begin{equation}
\lambda_h({\bf q})
\equiv
-\frac{\zeta_1 q_x^2+\zeta_2 q_y^2}{4\eta' |{\bf q}|}
+(\nu_1 q_x^2+\nu_2 q_y^2)
+\frac{K |{\bf q}|^3}{4\eta'}.\label{lamh_supple}
\end{equation}
We focus on the regime where pattern formation is possible, namely when the surface is stable at the largest length scales (\(\zeta_1,\zeta_2<0\)) but unstable at intermediate scales due to one or both of \(\nu_1,\nu_2<0\). For sufficiently small \(|{\bf q}|\), the \(\zeta_i\) terms dominate, making \(\lambda_h({\bf q})>0\) and thereby stabilizing the surface. As \(|{\bf q}|\) increases, the destabilizing \(\nu_i\) terms become important and can drive \(\lambda_h({\bf q})\) negative, leading to an instability. At still larger wave numbers, the free-energy contribution proportional to \(K |{\bf q}|^3\) dominates, rendering \(\lambda_h({\bf q})\) positive once again. Thus, the instability is confined to an intermediate range of wave numbers, giving rise to patterned states. See Fig.~\ref{lam_vs_q_diag}.
The onset of pattern formation and the corresponding preferred wavevector \({\bf q}_c\) can be determined from the extrema of the linear growth rate \(\lambda_h\). To this end, we set \(q_x=q\cos\theta\) and \(q_y=q\sin\theta\) in Eq.~(\ref{lamh_supple}) and impose the conditions
\[
\left.\frac{\partial \lambda_h}{\partial q}\right|_{q=q_c,\theta=\theta_c}=0,
\qquad
\left.\frac{\partial \lambda_h}{\partial \theta}\right|_{q=q_c,\theta=\theta_c}=0.
\]
Substituting \(q_x=q\cos\theta\) and \(q_y=q\sin\theta\) into Eq.~(\ref{lamh_supple}), we obtain
\begin{eqnarray}
\lambda_h(q)
&=&
-\frac{1}{4\eta' q}
q^2\left(\zeta_1\cos^2\theta+\zeta_2\sin^2\theta\right)
+q^2\left(\nu_1\cos^2\theta+\nu_2\sin^2\theta\right)
+\frac{Kq^3}{4\eta'}
\\
&=&
-\frac{\zeta(\theta)q}{4\eta'}
+\nu(\theta)q^2
+\frac{Kq^3}{4\eta'}
\\
&=&
\frac{Kq^3}{4\eta'}
+\nu(\theta)q^2
-\frac{\zeta(\theta)q}{4\eta'}.
\end{eqnarray}
Here, $\nu(\theta)=\nu_1\cos^2\theta+\nu_2\sin^2\theta$ and $\zeta(\theta)=\zeta_1\cos^2\theta+\zeta_2\sin^2\theta$. The linear theory is given by $\partial h/\partial t=-\lambda_h(q)\,h$ and the fastest growing mode corresponds to the most negative value of
\(\lambda_h(q)\), obtained from
\[
\left.
\frac{\partial \lambda_h}{\partial q}
\right|_{q=q_c,\theta=\theta_c}
=0.
\]
This gives
\[
\frac{3Kq_c^2}{4\eta'}
+
2\nu(\theta_c)\,q_c
-
\frac{\zeta(\theta_c)}{4\eta'}
=0,
\]
and hence
\begin{eqnarray}
q_c
&=&
\frac{-2\nu(\theta_c)
\pm\sqrt{4\nu^2(\theta_c)
+4\left(\frac{3K}{4\eta'}\right)
\left(\frac{\zeta(\theta_c)}{4\eta'}\right)}}
{2\left(\frac{3K}{4\eta'}\right)}
\\
&=&
\frac{2\eta'}{3K}
\left[
-2\nu(\theta_c)
+\sqrt{
4\nu^2(\theta_c)
+\frac{3K}{4\eta'^2}\zeta(\theta_c)
}
\right].
\label{qc-general}
\end{eqnarray}
 Here, we retain only the larger root of \(q_c\), since it corresponds to the dominant contribution to the growth rate, \(\exp[\nu(\theta_c)q_c^2(\theta_c)t]\). Next, we consider
\[
\left.
\frac{\partial \lambda_h}{\partial \theta}
\right|_{q=q_c,\theta=\theta_c}
=0.
\]
This gives
\[
\sin 2\theta_c
\left[
q_c(\nu_2-\nu_1)
-\frac{\zeta_2-\zeta_1}{4\eta'}
\right]
=0.
\]
Since $q_c=\frac{\zeta_2-\zeta_1}{4\eta'(\nu_2-\nu_1)}$ is unphysical, we only have, either
$\sin\theta_c=0$ or $\cos\theta_c=0,$ corresponding to modes along the \(x\) and \(y\) directions, respectively. Using these conditions together with Eq.~(\ref{qc-general}), we obtain two possible values of the preferred wavevector,
\begin{align}
q_c^x
=
\frac{2\eta'}{3K}
\left[
-2\nu_1
+\sqrt{
4\nu_1^2
+\frac{3K}{4\eta'^2}\zeta_1
}
\right],\\
q_c^y
=
\frac{2\eta'}{3K}
\left[
-2\nu_2
+\sqrt{
4\nu_2^2
+\frac{3K}{4\eta'^2}\zeta_2
}
\right];
\end{align}
see the main text.

\section{Direct numerical simulations}

\subsection{Numerical Method and Simulation Details}

Following Ref.~\cite{euler}, we numerically integrate Eq.~(\ref{basic-eq}) of the main text using a pseudo-spectral scheme with the (2/3) de-aliasing rule. where the linear term is treated exactly through an exponential propagator, while the nonlinear and noise terms are advanced using an Euler discretization~\cite{euler}, in the different linearly unstable cases to see if steady patterns are formed. Simulations are performed on a square lattice with lattice spacing $(a_0=1)$. The initial height profile was generated using uniformly distributed random numbers in the range \([-0.50,0.50]\). After the model attains the steady patterned phase, both the structure of the patterns and the associated height fluctuations remain stationary in time (see below). The resulting pattern structures remain unchanged for different random initializations with all other parameters fixed (see below). Other parameter values are listed in the figure captions.

\subsection{Anisotropic Pattern Formation}

We now discuss the anisotropic case in more detail. We first consider the case in which both \(\nu_1,\nu_2<0\) and \(\zeta_1,\zeta_2<0\), with \(\nu_1\neq\nu_2\) and \(\zeta_1\neq\zeta_2\). The corresponding patterns are shown in Fig.~\ref{iso_diag}(c,d,e) of the main text. Here, we provide a qualitative discussion of the physical mechanism underlying the observed pattern formation. Keeping the parameters \(\nu_1=-12\), \(\nu_2=-4\), \(\zeta_2=-0.02\) fixed, we vary \(\zeta_1\) to examine its effect on the periodicity.  Fig.~\ref{iso_diag}(c) of the main text shows a steady-state snapshot of the height profile for the parameter \(\zeta_1=-0.1\) corresponding to \(q_c^x\approx0.57\) and \(q_c^y\approx0.17\). The model selects the dominant mode, namely \(q_c^x\), leading to the formation of stripe-like patterns with periodicity along the \(x\) direction. These results are obtained for a small positive value of \(\lambda=0.001\). A qualitative explanation of this behavior originates from the competition between the destabilizing \(\nu_i\) terms and the stabilizing \(\zeta_i\) terms. In particular, the \(\zeta_1\) term competes with the destabilizing effect of the \(\nu_1\) term, while a similar competition occurs between the \(\zeta_2\) and \(\nu_2\) terms. Since \(|\nu_2|<|\nu_1|\), the combined effect of the \(\zeta_2\) term and the nonlinear \(\lambda\) term suppresses the instability associated with \(\nu_2\). Consequently, the model favors the mode \(q_c^x\), resulting in patterns aligned parallel to the \(y\) axis, i.e., with periodicity along the \(x\) direction. Fig.~\ref{iso_diag}(d) shows a steady-state snapshot for \(\zeta_1=-0.2\), corresponding to \(q_c^x\approx0.49\), while \(q_c^y\) remains unchanged. As \(q_c^x\) decreases, the periodicity increases, leading to a reduction in the number of stripes. Fig.~\ref{iso_diag}(e) shows a steady-state snapshot for \(\zeta_1=-0.4\), for which \(q_c^x\) becomes imaginary. Consequently, only the mode \(q_c^y\approx0.17\) survives, leading to patterns aligned parallel to the \(x\) axis with periodicity along the \(y\) direction. Intuitively, the \(\zeta_1\) term becomes strong enough to suppress the instability induced by \(\nu_1\), resulting in a change in both the periodicity and orientation of the stripes.

\begin{figure*}[]
\includegraphics[width=\textwidth]{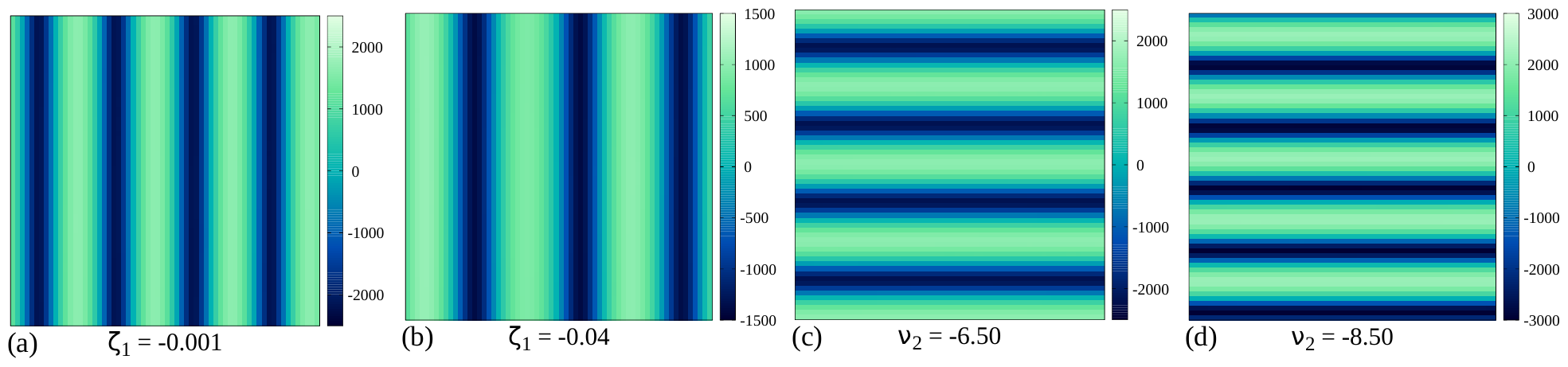}
\caption{Snapshots of the steady-state height profiles (\(L=64\)) in the patterned phase. Panels (a) and (b) correspond to \(\nu_1=-6.50\), \(\nu_2=1.50\), and \(\zeta_2=-0.001\), where varying \(\zeta_1\) changes the pattern periodicity. Panels (c) and (d) correspond to \(\nu_1=1.50\) and \(\zeta_1=\zeta_2=-0.001\), where varying \(\nu_2\) modifies the periodicity. In all cases, the remaining parameters are fixed at \(\eta'=0.02\), \(K=1\), \(\lambda=0.005\), \(D_h=0.10\), and \(\delta t=0.001\).}
\label{aniso_diag}
\end{figure*}

Next we consider the anisotropic case with \(\zeta_1,\zeta_2<0\), but one of \(\nu_1,\nu_2>0\), making the model unstable only along a particular direction. Fig.~\ref{aniso_diag}(a) and Fig.~\ref{aniso_diag}(b) correspond to \(\nu_2=1.50\), \(\zeta_2=-0.001\), and \(\nu_1=-6.50\), for which the model is stable along the \(y\) direction but unstable along the \(x\) direction. Consequently, only \(q_c^x\) exists. As \(\zeta_1\) changes from \(-0.001\) (\(q_c^x\approx0.34\)) to \(-0.04\) (\(q_c^x\approx0.29\)), the periodicity of the pattern changes. Fig.~\ref{aniso_diag}(c) and Fig.~\ref{aniso_diag}(d) show the complementary case where the model is unstable along the \(y\) direction. Here, the periodicity changes as \(\nu_2\) varies from \(-6.50\) (\(q_c^y\approx0.34\)) to \(-8.50\) (\(q_c^y\approx0.44\)).

\subsection{Additional Numerical Results}

\begin{figure*}[]
 \includegraphics[width=\textwidth]{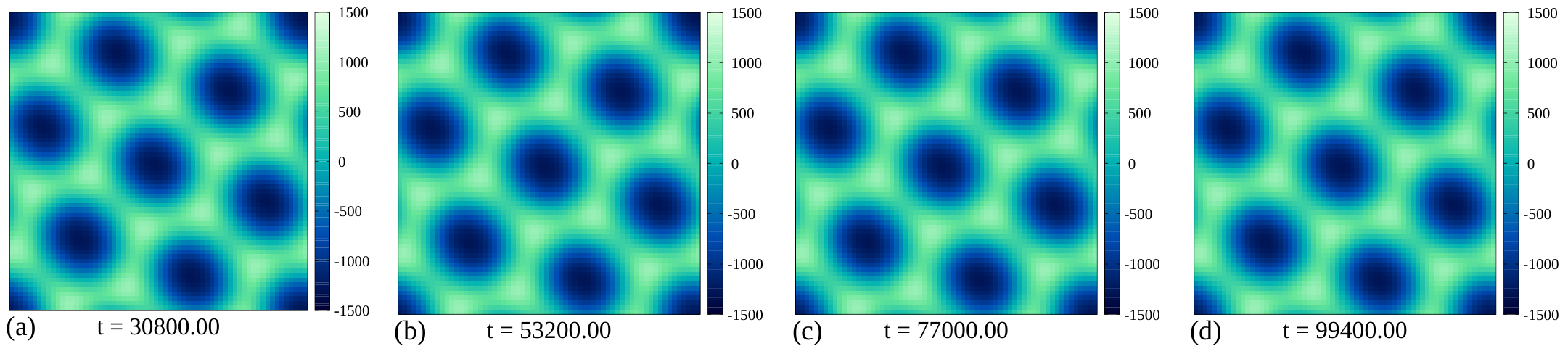}
\caption{Snapshots of the height profiles (\(L=64\)) in the isotropic patterned phase at different times after the steady state is attained. The persistence of the pattern demonstrates that both the morphology and the height fluctuations remain unchanged in time. The parameters are \(\nu_1=\nu_2=-6.50\), \(\zeta_1=\zeta_2=-0.063\), \(\eta'=0.02\), \(K=1\), \(\lambda=-0.005\), \(D_h=0.001\), and \(\delta t=0.01\).
}\label{supple_snap_diff_time}
\end{figure*}

Figure~\ref{supple_snap_diff_time} displays snapshots of the height field in the isotropic patterned phase at different times. The invariance of both the pattern morphology and the magnitude of the height fluctuations over time demonstrates the stability of the patterned state and confirms that the system has attained a stationary steady state.

\begin{figure*}[]
 \includegraphics[width=\textwidth]{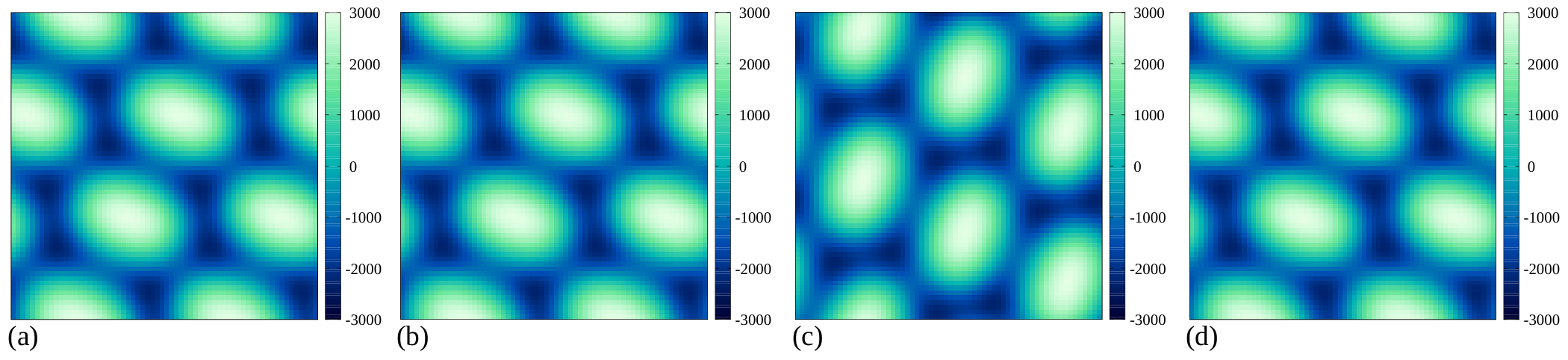}
\caption{Snapshots of the height profiles (\(L=64\)) in the isotropic patterned phase after the steady state is reached for different initial conditions at \(t=0\): (a) heights drawn from a uniform distribution in \([-0.50,0.50]\), (b) heights drawn from a uniform distribution in \([0.50,1.0]\), (c) heights drawn from a uniform distribution in \([-1.0,-0.50]\), and (d) a flat interface with \(h=0\) at every site. The identical steady-state morphology in all cases demonstrates that the patterned phase is robust against the choice of initial condition. The parameters are \(\nu_1=\nu_2=-6.50\), \(\zeta_1=\zeta_2=-0.0666\), \(\eta'=0.02\), \(K=1\), \(\lambda=0.0012\), \(D_h=0.001\), and \(\delta t=0.01\).
}\label{supple_diff_height_ini}
\end{figure*}

Figure~\ref{supple_diff_height_ini} shows steady-state snapshots of the isotropic patterned phase obtained from a variety of initial height configurations. Despite the markedly different starting conditions, all realizations evolve to the similar patterned state, indicating that the steady-state morphology is insensitive to the initial configuration and is an intrinsic feature of the dynamics.

\begin{figure*}[]
 \includegraphics[width=\textwidth]{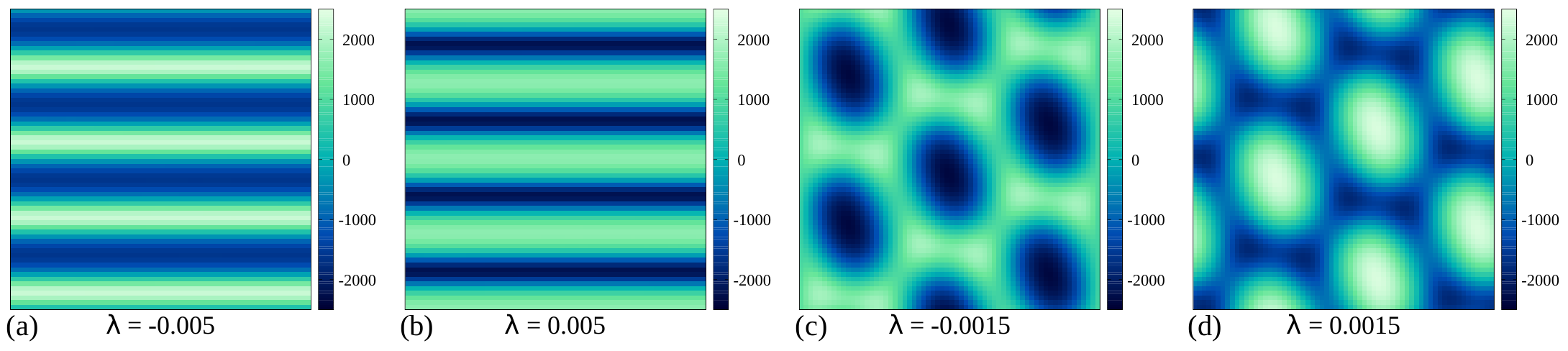}
\caption{Snapshots of the steady-state height profiles (\(L=64\)) for positive and negative values of \(\lambda\). Panels (a) and (b) correspond to the anisotropic patterned phase, while panels (c) and (d) correspond to the isotropic patterned phase. Reversing the sign of \(\lambda\) leaves the pattern morphology and periodicity unchanged but flips the height contrast, interchanging valleys and peaks (dark and bright regions). Parameters for (a) and (b): \(\nu_1=1.50\), \(\nu_2=-6.50\), \(\zeta_1=\zeta_2=-0.001\), \(D_h=0.10\), and \(\delta t=0.001\). Parameters for (c) and (d): \(\nu_1=\nu_2=-6.50\), \(\zeta_1=\zeta_2=-0.0666\), \(D_h=0.001\), and \(\delta t=0.01\). In all cases, \(\eta'=0.02\) and \(K=1\).
}\label{supple_pos_neg_lam}
\end{figure*}

Figure~\ref{supple_pos_neg_lam} presents steady-state height profiles for both positive and negative values of $(\lambda)$ in the anisotropic and isotropic patterned phases. Remarkably, reversing the sign of $(\lambda)$ leaves the characteristic length scale, orientation, and overall morphology of the patterns unchanged, while exchanging peaks and valleys of the interface. Consequently, regions that appear elevated for $(\lambda>0)$ become depressed for $(\lambda<0)$, and vice versa. This behavior reflects the role of the KPZ nonlinearity in breaking the inversion symmetry $(h\rightarrow -h)$; changing the sign of $(\lambda)$ effectively reverses the preferred direction of growth without altering the underlying instability responsible for pattern formation. This demonstrates that $\lambda$ primarily controls the polarity of the patterned interface, whereas the wavelength and spatial organization of the patterns are governed by the linear instability.

\begin{figure*}[t]
 \includegraphics[width=\textwidth]{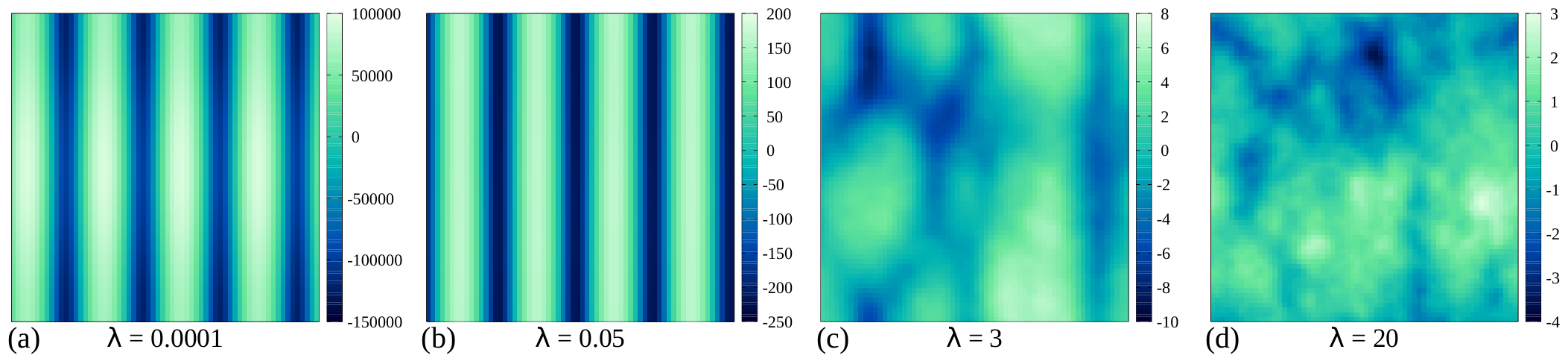}
\caption{Snapshots of the steady-state height profiles (\(L=64\)) in the anisotropic patterned phase for different values of \(\lambda\). Increasing \(|\lambda|\) strongly suppresses the height fluctuations, reducing the amplitude and visibility of the stripe patterns. For sufficiently large \(|\lambda|\), the patterned state is effectively destroyed. The parameters are \(\nu_1=-6.50\), \(\nu_2=1.50\), \(\zeta_1=\zeta_2=-0.001\), \(D_h=0.10\), \(\eta'=0.02\), \(K=1\), and \(\delta t=0.001\).
}\label{supple_aniso_vary_lam}
\end{figure*}

\begin{figure*}[]
 \includegraphics[width=\textwidth]{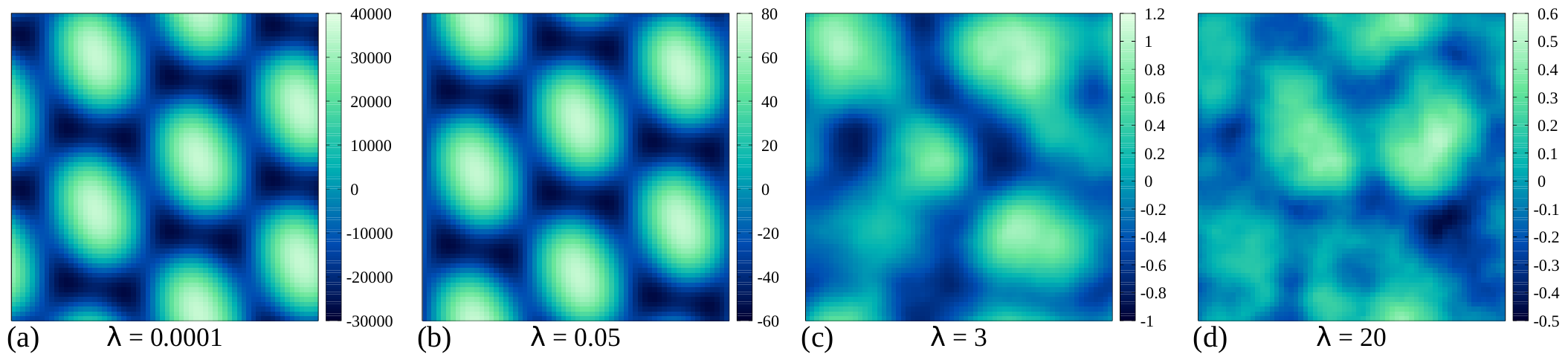}
\caption{Snapshots of the steady-state height profiles (\(L=64\)) in the isotropic patterned phase for different values of \(\lambda\). As \(|\lambda|\) increases, the height fluctuations are strongly suppressed and the amplitude of the isotropic mound-like structures decreases. For sufficiently large \(|\lambda|\), the patterned state disappears altogether. The parameters are \(\nu_1=\nu_2=-6.50\), \(\zeta_1=\zeta_2=-0.0666\), \(D_h=0.001\), \(\eta'=0.02\), \(K=1\), and \(\delta t=0.001\).
}\label{supple_iso_vary_lam}
\end{figure*}

Finally, Fig.~\ref{supple_aniso_vary_lam} and Fig.~\ref{supple_iso_vary_lam} illustrate the evolution of the steady-state height profiles with increasing \(|\lambda|\) for the anisotropic and isotropic cases, respectively. As \(|\lambda|\) increases, the amplitude of the height fluctuations is progressively reduced and the patterned structures become increasingly less pronounced, eventually disappearing altogether. This behavior can be understood from the stabilizing effect of the KPZ nonlinearity: larger values of \(|\lambda|\) generate stronger positive corrections to the effective \(\nu_i\), thereby suppressing the finite-wavelength instability responsible for pattern formation. Consequently, the system is driven away from the patterned phase toward a more stable, rough state.


\end{document}